# Annotation of *Tribolium* nuclear receptors reveals an evolutionary overacceleration of a network controlling the ecdysone cascade

François Bonneton[1,3]*, Arnaud Chaumot[2,3], Vincent Laudet[1]

[1] Université de Lyon, Ecole Normale Supérieure de Lyon, IGFL, CNRS UMR5161, INRA UMR1237, IFR128, 46 Allée d'Italie, 69364 Lyon Cedex 07, France

[2] CEMAGREF, Laboratoire d'écotoxicologie, 3bis quai Chauveau - CP 220 69336 LYON Cedex 09, France

[3] These authors contributed equally to this work

* Corresponding author. françois.bonneton@ens-lyon.fr (F. Bonneton)
fax: (33) 04 72 72 89 92

**Abstract**

The *Tribolium* genome contains 21 nuclear receptors, representing all of the six known subfamilies. When compared to other species, this first complete set for a Coleoptera reveals a strong conservation of the number and identity of nuclear receptors in holometabolous insects. Two novelties are observed: the atypical NR0 gene *knirps* is present only in brachyceran flies, while the NR2E6 gene is found only in *Tribolium* and in *Apis*. Using a quantitative analysis of the evolutionary rate, we discovered that nuclear receptors could be divided into two groups. In one group of 13 proteins, the rates follow the trend of the Mecopterida genome-wide acceleration. In a second group of five nuclear receptors, all acting together at the top of the ecdysone cascade, we observed an overacceleration of the evolutionary rate during the early divergence of Mecopterida. We thus extended our analysis to the twelve classic ecdysone transcriptional regulators and found that six of them (ECR, USP, HR3, E75, HR4 and Kr-h1) underwent an overacceleration at the base of the Mecopterida lineage. By contrast, E74, E93, BR, HR39, FTZ-F1 and E78 do not show this divergence. We suggest that coevolution occurred within a network of regulators that control the ecdysone cascade. The advent of *Tribolium* as a powerful model should allow a better understanding of this evolution.

# 1. Introduction

The recent burst of hexapod's genome projects has already provided two novel and major results concerning the evolution of holometabolous insects (Savard *et al.*, 2006a; Savard *et al.*, 2006b; Zdobnov and Bork, 2007). First, contrary to the most widely accepted hypothesis, Hymenoptera are basal to the other main holometabolous orders, Coleoptera, Diptera and Lepidoptera. Previous phylogenies, obtained with morphological and molecular markers (rRNA, mitochondrial DNA), were favouring a sister-group relationship between Hymenoptera and Mecopterida (Diptera+Lepidoptera), with Coleoptera as the basal group (Kristensen, 1999; Whitting, 2002). The new tree is fully resolved for these four large orders (>85% of hexapods species), although it lacks genomic data for seven smaller holometabolous orders. In that perspective, sequencing efforts for the Neuropterida superorder and for the enigmatic Strepsiptera would be highly valuable. The second important result is an acceleration of protein evolution in Mecopterida. Such an episodic change of rate had already been characterized for some genes in Diptera (Friedrich and Tautz, 1997), but the recent results show that this acceleration affected the whole genome of both Diptera and Lepidoptera (Savard *et al.*, 2006b; Zdobnov and Bork, 2007). Therefore, we can assume that an important evolutionary transition established a clear separation within holometabolous insects, between the monophyletic superorder Mecopterida and the non-Mecopterida species. This is very intriguing, because this molecular divergence is not obviously correlated with any major phenotypic change. The only morphological synapomorphies for Mecopterida are, for example, presence/absence of some specific muscles in adults or larvae (Kristensen, 1981; Whiting, 1998). A question is thus raised: what were the consequences of this acceleration for the developmental gene regulatory networks? Given the numerous interactions existing between proteins that control development, it is currently unclear how important functions can be maintained when their determining genetic elements are changing. Solving this issue requests to identify which part of a given network can change and how the different partners can coevolve. In view of the renewed landscape of holometabolous insect's phylogenomics, the Mecopterida acceleration appears as a case study to tackle these questions of the robustness and the adaptability of developmental regulatory networks during lineage specific events.

The regulatory networks that control the development of insects are largely composed of transcription factors and signalling proteins. Remarkably, one family of transcription factors, the nuclear receptors, can bypass the relatively slow and complex intracellular signalisation pathways. The transcriptional activity of these proteins usually depends on the binding of specific ligands to their ligand-binding domain. In animals, nuclear receptors are the only transcription factors (with the aryl hydrocarbon receptors) that are directly activated by small lipophilic ligands (hormones, fatty acids, gas) capable of going through the cell membrane. Nuclear receptors provide the organism with essential tools to respond rapidly, at the gene expression level, to environmental cues. The availability of the ligand coordinates, in time and space, the activity of these powerful gene regulators. They are thus involved in many physiological and developmental processes and, as a consequence, they are major targets of endocrine disruptors that are released in the environment by human activities (Henley and Korach, 2006). In insects, their role has been well characterised in various developmental processes such as: embryo segmentation, moulting, metamorphosis and eye morphogenesis (King-Jones and Thummel, 2005). They are very promising targets for the control of insect pests (Palli *et al.*, 2005). Interestingly, most nuclear receptors act as protein dimers and many of them can interact with each other in heterodimeric partnership, thus forming regulatory networks. The ecdysone regulatory cascade that controls metamorphosis in *Drosophila*, where 9 out of 18 nuclear receptors are involved, best illustrates these crosstalks.

Thanks to the sequencing of the *Tribolium* genome by the Baylor Human Genome Sequencing Center, we were able to identify the first complete set of nuclear receptors for a Coleopteran insect. This provides the opportunity for a phylogenetic analysis of these proteins encompassing the four major groups of holometabolous insects. Since we have described earlier the acceleration of the ecdysone receptor (ECR-USP) in Mecopterida (Bonneton *et al.*, 2003; Bonneton *et al.*, 2006), we ask here whether the other nuclear receptors acting in the ecdysone cascade were affected similarly. Our analysis suggests how the different partners of an essential developmental regulatory network can coevolve through a lineage specific acceleration.

**2. Materials and methods**

*2.1 Annotation and phylogenetic analysis of Nuclear Receptors*

We used the nuclear receptors sets of *Drosophila melanogaster* and *Apis mellifera* (Velarde *et al.*, 2006) to query the *Tribolium castaneum* genome (version 2.0). The same approach was used against Genbank in order to recover all the nuclear receptors protein sequences from the six other insect's species whose genome was available (Fig. 1). When a nuclear receptor was missing, or was too short for one species, nucleic acid sequences were retrieved and analysed with two gene prediction programs: Augustus (Stanke *et al.*, 2006) and Genescan (Burge and Karlin, 1997). When different isoforms were recovered, only the longest one including a DBD and a LBD was chosen for analysis. Predicted protein-coding sequences were aligned using SEAVIEW (Galtier *et al.*, 1996) and manual corrections were made, when possible, following the phylogenetic trees and by the structural data (FCP web tool: Garcia-Serna *et al*., 2006). All the *Tribolium castaneum* nuclear receptors genes could easily be identified (table 1). Note that this task was facilitated by the fact that TLL, EG, ECR and USP had been cloned prior to the sequencing of *Tribolium* genome (Schröder *et al*., 2000; Bucher *et al*., 2005; Iwema et al., submitted). By contrast, only 17 nuclear receptors were identified for *Bombyx*, among which only 12 are long enough to be included into the analysis. The vertebrates sequences, mainly retrieved from NuReBase, were used as outgroup (Ruau *et al*., 2004). Phylogenetic reconstruction was made with the BIONJ algorithm (Gascuel, 1997), an improvement of the Neighbor Joining method (Saitou and Nei, 1987), with Poisson correction for multiple substitutions, or with the maximum parsimony method, as implemented in Phylo_Win (Galtier *et al*., 1996). All positions with gaps were excluded from analyses.

*2.2 Quantitative analysis of nuclear receptors evolution*

We aligned all the available protein sequences of arthropods independently for each of the 18 nuclear receptors that possess a LBD and that are found in all insects. This excludes NR2E6, found only in *Apis* and *Tribolium*, and the proteins of the NR0 subfamily, which do not possess a LBD. The same procedure was applied for four other transcription factors (E74, BR, Kr-h1 and E93) involved in the control of the ecdysone pathway (Fig. 5). Alignments were automatically performed using ClustalW (Thompson *et al*., 1994) with manual correction in Seaview (Galtier *et al*., 1996). After removing of partial uninformative sequences, we considered only the sequences from holometabolous insects. All positions with gaps and misaligned regions were removed, resulting in alignments of protein regions mostly encompassed in the DBD and LBD domains. Only four species allowed to recover a good set of alignments allowing the comparison between all the 18 nuclear receptors: *Drosophila* *melanogaster*, *Aedes aegypti*, *Tribolium castaneum* and *Apis mellifera*. One exception is HR83, for which we had to use the sequence of the closely related mosquito *Anopheles gambiae* instead of the short sequence (85 amino-acids) of DBD identified from *Aedes aegypti* genome. We performed supplementary analysis with a set of five species, adding the sequences of *Bombyx mori* available for 12 of the 18 nuclear receptors.

The pattern of evolution of each nuclear receptor was determined by the calculation of the branch lengths of a phylogenetic tree gathering the four or five species, using a predefined unrooted topology (Fig. 4C; Supp. Fig. 2C). Branch lengths were estimated with maximum likelihood methods using the PAML program (Yang, 1997). Likelihood calculations were performed under the JTT amino acid substitution model (Jones *et al.*, 1992) plus rate heterogeneity between sites, estimated by a discrete gamma law with six categories (with the shape parameter as an additional free parameter). In order to check whether the estimation of the distances with a small number of species was robust enough and avoid a possible taxonomic bias, we constructed trees including all the arthropods sequences available for each protein. Then, we extracted the "subtrees" corresponding to the four or five species of reference. The comparison of the distances obtained with both sets of species revealed a very good linear correlation ($R^2 = 0.98$; p-value $< 10^{-15}$) with values closed to the equality (slope of the linear regression comprised between 0.9 and 1 for a linear model with a null intercept).

We adopted an approach used in morphometric analysis to compare, in a quantitative manner, the evolution of the nuclear receptors during the radiation of holometabolous insects. Using either the set of 18 trees established with four species, or the set of 12 trees with five species, we performed principal component analysis (PCA) to compare the patterns of evolution of the different nuclear receptors. The unrooted tree computed for each nuclear receptor was then considered as an « individual » that could be described by as many variables as branches: 5 variables for four species and seven variables for five species. We performed non-normed PCA using the package ade4 (Chessel *et al.*, 2004) of the R software (R Development Core Team, 2006). We displayed factorial maps (Fig. 4AB) by means of the biplot procedure, which allows visualising simultaneously the distribution of individuals and the correlation between variables and principal axes (Chessel *et al.*, 2004).

In order to discriminate groups of proteins with similar patterns of evolution, we performed hierarchical clustering analysis based on factorial coordinates following four distinct agglomeration methods (Ward's method, complete, single and average linkage method) as implemented in the package stat of the R software (R Development Core Team, 2006). Bold branches on figure 4D and supplemental figure 2D underline the clusters, which are found with all the four different hierarchical methods.

To compare the patterns of evolution observed for nuclear receptors with the global genomic trend during holometabolous radiation, insect phylogenomic trees from the literature, estimated with 64,134 aa (Savard *et al.*, 2006a) 705,502 aa or 336,069 aa (Zdobnov and Bork, 2007), were projected on the factorial maps established with trees computed for nuclear receptors (Fig. 4).

In addition, the same PCA procedure as in figure 4 was performed only with the phylogenetic trees of eight nuclear receptors (ECR, USP, E75, HR3, HR4, E78, HR39 and FTZ-F1) plus four other transcription factors (E74, Kr-h1, E93 and BR), forming a sample of twelve genes involved in the control of the ecdysone pathway (King-Jones and Thummel, 2005). For this analysis (Fig. 5), we did not comprise the length of the *Drosophila* branch among the variables of the PCA because of an atypical strong divergence of the *Drosophila* protein E93. Indeed, this long branch in the tree of E93 conceals the pattern observed if we exclude E93 from the analysis. Conversely, the reported PCA (Fig. 5) yields factorial maps closed to the scatter diagrams obtained without E93, and moreover the pattern disclosed on these diagrams is not affected whether we do not include the *Drosophila* branch in the variables of the PCA taking into account only the eleven remaining proteins (data not shown).

**3. Results**

*3.1 The genome of Tribolium castaneum contains 21 nuclear receptors*

Thanks to the conserved DBD and LBD domains, all the *Tribolium castaneum* nuclear receptors genes could be identified by blast searches on the available gene predictions (GLEAN, Genbank). The *Tribolium* genome contains 19 typical nuclear receptors, representing all of the six subfamilies described so far. An additional subset of two nuclear receptors lacking a LBD (subfamily NR0, group A) is also present (table 1 ; Fig. 1). Overall, as expected, the DBDs show high conservation (76-99%) while the LBDs are more divergent (26-94%) when compared to the *Drosophila* orthologs. The most divergent nuclear receptor is the insect's specific HR83 (NR2E5), of unknown function, while the most conserved is SVP (NR2F3), the ortholog of COUP-TF, an orphan receptor essential for metazoan development. Unlike the *Hox* genes, which are clustered on one single complex, we could map 18 nuclear receptors genes on seven of the ten linkage groups of *Tribolium* (table 1).

The identification of *Tribolium* nuclear receptors reveals the first complete set for a Coleoptera. It is therefore now possible to compare this family of proteins between all the major orders of holometabolous insects.

*3.2 Comparative analysis of nuclear receptors in insects*

The set of 21 *Tribolium* nuclear receptors is very close to the set of other insects, which range from 20 in *Aedes aegypti* to 22 in honeybee (Fig. 1). The novelties are restricted to two genes, which are present in some groups of species and not in others.

First, the gap gene *knirps* is found only in brachyceran flies (*Drosophila* and *Musca domestica*), but not in mosquitoes or any other holometabolous insect (fig 1; fig 2A). All species studied so far (including *Tribolium* and *Apis*) possess at least two NR0 genes, suggesting that the duplication, which produced *eagle* and *knirps-like*, arose early during insect evolution. After duplication, *knirps* diverged rapidly from its paralog, which is probably *knirps-like*, as suggested by the chromosomal positions and the strong conservation of function during development between these two genes (Rothe *et al.*, 1992; Lunde *et al.*, 2003). All the atypical NR0 genes are located on a single chromosome in the genome of *Tribolium* (LG3), *Drosophila* (3L, 77CE-78E) and *Anopheles* (3L, 38B) (supplementary table).

By contrast, the NR2E6 gene was specifically lost in Diptera, and maybe in Lepidoptera as well. This gene has no vertebrate homolog and it has been identified only in *Tribolium* and in *Apis*, with 67% of overall similarity between the proteins of these two species (97% for the DBD only). The NR2E group contains several insect specific nuclear receptor, such as *dissatisfaction* and HR83, together with genes that share clear homologs with vertebrates, such as *tailless* and HR51/PNR (NR2E3). The phylogenetic analysis of this group shows that NR2E6 is a new insect's specific nuclear receptor of the NR2E subgroup that is not significantly related to NR2E3 (Fig. 2B). In *Tribolium*, HR51 (NR2E3) is located on the linkage group 7 while NR2E6 is on the LG5. Therefore, we suggest using the nomenclature-based name NR2E6 for this gene, rather than the trivial name PNR-like, which were both proposed previously by Velarde *et al.* (2006).

Since both ECR and USP experienced a strong acceleration of evolutionary rate in Mecopterida (Bonneton *et al.*, 2006), we looked whether a similar trend could be observed for other nuclear receptors. As a first step for this test, we performed a simple phylogenetic analysis with each of the 18 nuclear receptors that possess a LBD and that are found in all insects. This excludes NR2E6 and the proteins of the NR0 subfamily. It is not our aim to provide here a full phylogeny of the whole family, but rather to use the trees to detect possible accelerations (Bonneton *et al*. 2003). If we consider only the proteins with orthologs available, at least, in Diptera, Lepidoptera, Coleoptera and Hymenoptera, the results reveal two kinds of topologies: either a well supported divergence of the Mecopterida (Diptera+Lepidoptera) branch, or not (table 2). On the first group we find E75, HR3, ECR, USP, HR78 and HR4. The DBD sequences are highly conserved (Supp. Fig. 1), while the Mecopterida specific differences are scattered along the LBD domain, which show low identity percentage (table 1). As an example, the figure 3 shows an unrooted tree of NR1D and NR1F proteins, which correspond to, respectively, E75/REV-ERB and HR3/ROR. Note that, in contradiction with the known phylogeny, *Tribolium* sequences are grouped with *Apis* and other non-Mecopterida orthologs. This aberrant topology is due to a long branch attraction, because of the acceleration of evolutionary rate in the Mecopterida lineage. Similar results were already described for ECR and USP (Bonneton et al., 2003). By contrast, HNF4, TLL, SVP, HR38, FTZ-F1 and HR39 have a much more conserved LBD sequence (table 1) and their respective tree do not show a significantly supported Mecopterida branch (table 2; see also figure 2B for an example of such a tree with TLL).

In conclusion, the main event that occurred during the evolution of nuclear receptors in holometabolous insects is probably the strong acceleration of some of its members in the Mecopterida lineage. In order to characterize further this spectacular divergence, we decided to perform a quantitative and comparative analysis of the rates of divergence.

*3.3 Overacceleration among nuclear receptors in Mecopterida*

The evolutionary pattern of nuclear receptors was analysed using a quantitative comparison of the divergence within a common set of holometabolous insect species. Branch lengths of phylogenetic tree for each nuclear receptor were first computed by maximum likelihood methods. Then, the computed trees were compared through a morphometric approach, by means of a principal component analysis (PCA). Here, we considered each tree as an « individual » harbouring a morphology with a specific size (the total length of the tree, *i.e.* the average number of substitutions per site that occurred during the evolution of holometabolous insects) and a specific form (the relative lengths of the branches, *i.e.* the divergence observed along each lineage). Despite the availability of the *Bombyx* genome, it was sometimes impossible to recover some suitable sequences for Lepidoptera. Therefore, we performed analyses with four species (all the 18 nuclear receptors) or with five species (only 12 proteins) (see materials and methods). Both studies reveal the same pattern on factorial maps (Fig. 4A,B; Supp. Fig.2A,B).

Nearly all variables (branch lengths) are correlated with the first principal axis that explains a large part of the variance: 57% with four species and 63% with five species (Fig. 4A; Supp. Fig.2A). This is consistent with a classical result in morphometry, where the first axis of the PCA translates the variation of the global size of individuals, in our case the variation in the total length of the phylogenetic trees (Jolicoeur and Mosimann, 1960). In other words, the first axis ranks the nuclear receptors according to the average amount of substitutions per site. If we consider that the selected sites for each of the nuclear receptors constitute representative and comparable samples of each gene (regions encompassed in the DBD and LBD domains), then the first axis distributes nuclear receptors along a gradient from genes with most constrained evolution (for example: SVP, FTZ-F1, HR39, HR38) to genes with higher rates of evolution in holometabolous insects (for example: USP, HR78, ERR, E78, HR83).

The second principal axis supports a large part of the remaining variance: 38% with four species and 61% with five species (Fig. 4B; Supp. Fig.2B). This remaining variance translates the diversity of the evolutionary patterns observed between nuclear receptors, if we exclude the heterogeneity of their global evolution rates viewed on the first principal axis. Strikingly, the second axis is highly correlated with only one variable in both cases: the length of the "Mecopterida-Diptera" branch for the four species trees or the length of the "Mecopterida" branch for the five species trees. Importantly, this variable is poorly correlated with PCA axis 1, showing that the global evolutionary rate of each protein does not explain the variability of the divergence along this branch. Furthermore, the second principal axis is remarkably supported by the existence of a highly discriminated group of five nuclear receptors : HR4, E75, USP, ECR and HR3 (Fig. 4B; Supp. Fig.2B), which show a longer Mecopterida branch, when compared to other nuclear receptors. This group of five proteins with a strong divergence along the Mecopterida stem branch is also clearly revealed by hierarchical clustering analysis based on factorial coordinates (Fig. 4D; Supp. Fig.2D).

Since a genome-wide acceleration occurring along the "Mecopterida" and "Diptera" branches was reported recently for housekeeping genes (Savard *et al*. 2006a ; Savard *et al*. 2006b) and for a larger sample of single-copy orthologs (Zdobnov and Bork, 2007), we examined the specificity of the evolutionary acceleration of these five nuclear receptors by projecting published insect's phylogenomic trees onto the factorial maps established for nuclear receptors (Fig. 4AB; Supp. Fig.2AB). Considering that the projected points are not clustered with the five discriminated proteins and that they are scattered within the group of other nuclear receptors, we conclude that the acceleration affecting HR4, E75, USP, ECR and HR3 constitutes an additional event to the global genomic trend. All the other nuclear receptors followed the trend of evolution that characterise the Mecopterida divergence. Interestingly, we can also notice that the projection of the phylogenomic tree estimated with housekeeping proteins (Savard et al, 2006b) is placed on the left of the factorial map, close to more constrained genes (▲ in Fig. 4A). Since housekeeping proteins are assumed to be under strong selective constraints, this result is consistent with the interpretation of the first principal axis.

Thus, HR4, E75, USP, ECR and HR3 underwent an overacceleration of evolutionary rate during the emergence of the Mecopterida clade, which did not affect the other nuclear receptors.

*3.4 Coevolution at the top of the ecdysone regulatory cascade*

Remarkably, the five overaccelerated nuclear receptors act together in the upstream part of the ecdysone regulatory cascade that triggers *Drosophila* metamorphosis (King-Jones and Thummel, 2005). ECR and USP constitute the heterodimeric ecdysone receptor, E75 is a primary early response gene and HR3 and HR4 are early late genes. However, other nuclear receptors acting early during this hormonal response, such as E78, HR39 or FTZ-F1, do not show this overacceleration (Fig. 4). It seems, therefore, that only some part of the upstream ecdysone regulatory network may have evolved rapidly in Mecopterida. In order to test this hypothesis, we completed a principal component analysis with twelve transcription factors known to regulate the top of this cascade: the eight nuclear receptors described above, plus E74 (Ecdysone induced protein 74EF), BR (Broad), E93 (Ecdysone induced protein 93F) and Kr-h1 (Kruppel-homolog 1). The factorial map discloses a clear separation between two groups of proteins (Fig. 5). We find the same cluster of five nuclear receptors, plus Kr-h1, with the overacceleration along the "Mecopterida-Diptera" branch. The other group contains E74, E93, BR and the three nuclear receptors HR39, FTZ-F1 and E78.

This result shows that six out of the twelve classic transcriptional regulators known to act at the top of the ecdysone pathway underwent an overacceleration in Mecopterida. Consequently, we can assume that coevolution probably occurred between a sub-network of overaccelerated nuclear receptors that control the ecdysone regulatory cascade (Fig. 6).

## 4. Discussion

*4.1 The set of nuclear receptors is conserved in holometabolous insects*

The set of nuclear receptor genes in holometabolous insects ranges from 20 in *Aedes aegypti* to 22 in honeybee (Fig. 1). The evolution of this metazoan protein family is complex, with many variations (duplications, losses) around a common theme of six subfamilies (Bertrand *et al*., 2004). Unlike nematodes, where the genome of *Caenorhabditis elegans* and *Caenorhabditis briggsae* contain 283 and 268 nuclear receptors, respectively (Stein *et al*., 2003), the monophyletic group of holometabolous insects did not experience a lineage-specific expansion within the nuclear receptors family. If more genomic data are needed to understand the surprising diversity observed in ecdysozoans, it is now clear that there is a strong conservation of the number and identity of nuclear receptors in holometabolous insects.

One novelty is the presence of the NR2E6 gene in *Tribolium* and in *Apis* but not in Diptera. In honeybee, the transcripts of this gene were found in the brain and in the eye of pupa and adult, a pattern of expression which is reminiscent of the retina-specific pattern of PNR, the human homolog of HR51 (Velarde *et al*., 2006). Interestingly, all the members of the NR2E group apparently share a primary function in the developing nervous system (Laudet and Bonneton, 2005). The beetle genome, like the honeybee, is less derived than the Diptera genome and contains more ancestral genes. It is possible that NR2E6 is one of these ancestral genes that will eventually be identified in other arthropods. The question of its origin remains open, since it is absent in vertebrates and in nematodes.

The fact that the two model organisms, *Drosophila* and *Tribolium*, have very similar sets of nuclear receptors is very promising for the understanding of this family in insects. Indeed, it means that genetic and physiologic studies based on both species will complement each other and should have general implications for other holometabolous insects. However, homologous genes can give different proteins, because of divergent evolutionary rates that can occur even in the absence of gene duplication and loss. In that respect, USP, HR78, ERR, E78 and HR83 seem to be less constrained, with higher rates of evolution in holometabolous insects, when compared to SVP, FTZ-F1, HR39 or HR38 (Fig. 4A). Such divergences must be taken into account for future comparisons, as evidenced by our work showing that, if USP has a large liganded pocket in *Drosophila* and in the moth *Heliothis*, it is an orphan receptor with no ligand binding pocket in *Tribolium* (Clayton *et al*., 2001; Billas *et al*., 2001; Iwema et al., submitted). The opportunity to analyse *Drosophila* and *Tribolium* at the same time should reveal more about such fundamental differences between nuclear receptors. *Tribolium* is particularly favoured as a model, as its development is more representative of the early holometabolous insects.

*4.2 Nuclear receptors display two modes of evolutionary rate in holometabolous insects*

We have previously shown that two nuclear receptors, ECR and USP, underwent an acceleration of evolutionary rate in Mecopterida (Bonneton *et al*., 2003; Bonneton *et al*., 2006). Both proteins heterodimerise to constitute the ecdysone receptor in insects as well as in crabs and ticks (Henrich, 2005). Therefore, they act together at the top of an essential hormonal pathway that controls developmental timing and metabolism in arthropods. Since both proteins are involved in so many vital interactions, such a divergence must require coevolution of their other partners. Actually, it was revealed recently that a genome-wide acceleration occurred along the Mecopterida branch (Savard *et al*. 2006a; Savard *et al*. 2006b; Zdobnov and Bork, 2007). Thus, it was possible that other nuclear receptors experienced the same evolution. This possibility was tested by a quantitative analysis of the evolutionary rate, which revealed two important features of this family.

First, we have found that nuclear receptors show a "gradient" of average substitution rates during the radiation of holometabolous insects, from slow evolving proteins, whose structure and function are known to be highly conserved throughout animals, such as SVP/COUP-TF, HNF4, or HR38/NURR1, to fast evolving proteins, such as HR83 or E78. Assuming that housekeeping genes, which by definition are expressed in all cells and at all times, are under strong purifying selection, the comparison of evolution patterns between nuclear receptors and housekeeping genes (Savard *et al*. 2006a ; Savard *et al*. 2006b) or genes from larger genomic samples (Zdobnov and Bork, 2007) leads us to conclude that the majority of nuclear receptors underwent high selective pressure in insects. Only HR83, E78, ERR, HR78 and USP show a likely more relaxed evolution than housekeeping genes.

Second, our results show that nuclear receptors can be divided into two groups, according to their rate of evolution during the early divergence of the Mecopterida clade. In one group of 13 proteins, the rate is similar to the Mecopterida genome acceleration (Savard *et al*. 2006b). In a second group of five nuclear receptors (ECR, USP, HR3, E75 and HR4), we observe an overacceleration of the evolutionary rate along the Mecopterida stem branch, which is suggestive of a release of selective pressure after the initial event of genome-wide acceleration. Notably, this putative release plays on the LBD, but not on the DBD, which structure and sequence remained very constrained in every receptor (Table 1). This overacceleration can be detected by a simple phylogenetic analysis, producing trees where the Mecopterida species are significantly separated from the other holometabolous species. The only exception is HR78, which shows a Mecopterida divergence on the trees, despite a lack of overacceleration. In that case, the aberrant topology is likely due to the extreme divergence of the *Bombyx mori* sequence (Fig. 4)(Hirai, 2002). The PCA method is thus more reliable to detect such events, especially if the taxonomic sample is small and not fully representative of the phylogeny.

We can now conclude that the acceleration of ECR and USP observed initially is in fact an overacceleration, which concerns three other nuclear receptors as well. This overacceleration occurred during the diversification of Mecopterida, approximately 280-300 million years ago (early Permian). In any rigour, the fact that this divergence is Mecopterida specific requires analysis of sequences from all groups of this superorder, not only Diptera and Lepidoptera, but also Trichoptera, Mecoptera and Siphonaptera. This evidence has been provided for ECR and for USP (Bonneton *et al.*, 2006).

*4.3 Overacceleration at the top of the ecdysone cascade in Mecopterida*

If we consider the five overaccelerated nuclear receptors (E75, HR3, ECR, USP and HR4), they share an obvious common characteristic: they all act at the top of the ecdysone cascade that triggers metamorphosis. Most effects of ecdysone are mediated through the heterodimeric ECR-USP receptor that directly regulates the transcriptional activity of the three other nuclear receptors. E75 (Eip75B, the classic puff at E75B) is induced as a primary early response gene, while HR3 and HR4 are induced as early late genes (King-Jones and Thummel, 2005). HR3 is induced after puparium formation, represses early genes and is a direct activator of the prepupal regulator FTZ-F1. In *Drosophila*, as well as in *Bombyx*, E75 acts as a repressor of HR3, through direct heterodimerisation (White 1997; Swevers 2002; Hiruma 2004; Palanker *et al.*, 2006). HR4 acts with HR3 in the regulation of target genes, including FTZ-F1 (King-Jones *et al.*, 2005). Therefore, cross-regulatory interactions between E75, HR3 and HR4 converge on FTZ-F1 to discriminate between the ecdysone responses of the first (puparium) and second (pupation) hormonal peaks that initiate the metamorphosis process. All these results show that the five overaccelerated nuclear receptors are important players of the same regulatory network.

Several other proteins are known to act early in the ecdysone pathway. Among these classic regulators are the nuclear receptors E78 and HR39, as well as different transcription factors, such as: E74, BR, E93 and Kr-h1. Our results show that these genes did not experienced the Mecopterida overacceleration. One exception is Kr-h1, an ecdysone-regulated gene encoding a zinc-finger protein, which modulates the prepupal response (Pecasse *et al.*, 2000; Beck *et al.*, 2005). Unfortunately, nothing is known about its possible contacts with other key proteins of the ecdysone pathway. We hypothesize that the upstream part of the ecdysone cascade includes at least one network of closely interacting proteins that might be physically independent of the other regulators. This modularity would explain the coevolution of the five nuclear receptors that act together. If one member of the network suddenly accumulates mutations, then a parallel overacceleration of its partners would help to maintain their interactions. The rate of evolution is higher when the connecting proteins have transient interactions, which is the case for nuclear receptors (Pal *et al.*, 2006). In such a scenario, the interface domains would be the main targets of molecular adaptation. Interestingly, by comparing *Drosophila* and *Tribolium*, we found that the ecdysone binding ability of ECR has not changed during this evolution. However, the heterodimerisation surface between ECR and USP has accumulated changes, therefore creating a new interface (unpublished results). In the same line of idea, it would be very interesting to compare the evolution of the dimerisation contacts that occur between HR3 and E75. If physical interactions often induce coevolution, then coevolution can help to detect new interactions. Indeed, different methods are using coevolution between proteins, domains, or even between amino acids to predict biological networks (Pazos *et al.*, 1997; Lichtarge *et al.*, 2003; Fraser *et al.*, 2004).

The patterns of evolution among nuclear receptors are in fact not always so straightforward. For example, USP also heterodimerise with HR38, SVP and HR78 (Baker *et al.*, 2003; Miura *et al.*, 2002; Zhu *et al.*, 2003; Hirai *et al.*, 2002). However, none of these proteins shows the Mecopterida overacceleration. Some of these interactions are even conserved in vertebrates, such as RXR (USP) with NURR1 and NGFIB (HR38). In such cases, it is possible that coevolution concerns only few amino acids, resulting in an undetectable acceleration in our analysis of the whole LBD. Testing this possibility requires to determine the structure of the heterodimer, in order to map the putative accelerated residues.

*4.4 Maintenance of the ecdysone pathway*

If the proteins of a network controlling the ecdysone cascade have diverged, then what about the network itself? In other words, is the ecdysone response different between Mecopterida and other holometabolous insect's species? Most of the functional studies have been done using Diptera and Lepidoptera species, and the advent of *Tribolium* as a new model will allow filling this lack of data. However, all available evidences suggest that this vital hormonal control is well conserved among insects (Truman and Riddiford, 2002; Lafont *et* *al*., 2005). To cite only the most recent and compelling results, Xavier Bellés and his colleagues, in Barcelona, have shown, using the heterometabolous insect *Blattella germanica*, that the phenocopies of ECR, USP and HR3 genes mimic very closely the phenotype of the corresponding mutants in *Drosophila* (Cruz *et al*., 2006; Martin *et al*., 2006; Cruz *et al*., 2007). These *Blattella* proteins are more similar to their *Tribolium* homologs than to their Mecopterida homologs. Therefore, we can reasonably assume that, despite the acceleration of six major regulators acting at the top of the ecdysone cascade, the output of this pathway is very likely conserved among holometabolous and heterometabolous insects. This view is compatible with the coevolution hypothesis presented above: "*For things to remain the same, everything must change.*" (Tomasi di Lampedusa, 1958).

**Acknowledgments**

MENRT, Université de Lyon, CNRS and ENS funded this work.

**List of supplementary material**

Supplementary Table. Chromosomal location of insect's nuclear receptors.

Supplementary Figure 1. Alignment of the DBD protein sequences of the five overaccelerated nuclear receptors. Dots indicate identical amino acids, when compared to *Drosophila melanogaster*. The positions of cysteine residues of the zinc-finger are highlighted in grey.

Supplementary Figure 2. Principal component analysis of the evolution of nuclear receptors of holometabolous insects. A non-normed PCA was performed using the branch lengths of a predefined phylogenetic tree (C) computed for twelve nuclear receptors with identified ortholog sequences in *Drosophila melanogaster*, *Aedes aegypti*, *Bombyx mori*, *Tribolium castaneum* and *Apis mellifera*. On the PCA factorial maps (A, B), the seven variables (*i.e.* lengths of the seven branches called dmel, aaeg, Diptera, bmor, Mecopterida, tcas and amel on C) are symbolized by arrows and superimposed on the individuals (nuclear receptors). The plots display, either the first and second principal axes (A), or the second and third principal axes (B). Eigenvalues bar charts show, in black, the two axes used to draw each biplot. Clustering dendrogram (D) based on the position of nuclear receptors on the factorial map 2-3 (B) was computed following Ward's method. Bold branches underline the clusters, which are found using four different hierarchical methods (see materials and methods). The supplementary points on A and B correspond to the projections of insect's phylogenomic trees obtained with concatenated alignments of large numbers of genes: ▲ 33,809 aa (Savard *et al*., 2006b), ■ 336,069 aa and ● 705,502 aa (Zdobnov and Bork, 2007).

**Figure legends**

Figure 1. Nuclear receptors of holometabolous insects. Both the usual *Drosophila* name and the official nomenclature name of the proteins are given. For each nuclear receptor, a coloured box indicates its presence/absence in the genome for each of the seven species sequenced so far. The tree on the left shows the phylogeny of the species, with the Mecopterida indicated in red. The tree at the bottom indicates the putative relationships between the nuclear receptors (Bertrand *et al*., 2004). The nuclear receptors that experienced an overacceleration in Mecopterida are highlighted in red. Note that, for *Bombyx*, the current status of the genome sequence does not allow to determine the presence/absence of some NR2E genes, as symbolised with question marks and dotted-line boxes.

Figure 2. Phylogeny of the NR0 subfamily (A) and of the NR2E group (B) in insects. Unrooted trees were constructed using the Neighbour Joining method with the maximum length of sequence, resulting in 140 complete aligned sites for NR0 and 183 sites for NR2E.

Bootstrap values (Neighbour Joining/Maximum Parsimony) are indicated only for branches discussed in the text. The names of proteins and species are those indicated in figure 1. *Tribolium* nuclear receptors are highlighted with a black arrowhead (◄). Measure bar: differences per site.

Figure 3. Phylogeny of E75 and HR3. This unrooted tree was constructed using the Neighbour Joining method with the maximum length of sequence, resulting in 311 complete aligned sites. Bootstrap values (Neighbour Joining/Maximum Parsimony) are indicated only for branches discussed in the text. The names of proteins and species are those indicated in figure 1. *Tribolium* nuclear receptors are highlighted with a black arrowhead (◄). Measure bar: differences per site.

Figure 4. Principal component analysis of the evolution of nuclear receptors of holometabolous insects. A non-normed PCA was performed using the branch lengths of a predefined phylogenetic tree (C) computed for 18 nuclear receptors with identified ortholog sequences in Drosophila *melanogaster*, *Aedes aegypti*, *Tribolium castaneum* and *Apis mellifera*. On the PCA factorial maps (A, B), the five variables (*i.e.* lengths of the five branches called: dmel, aaeg, amel, tcas and Mecopterida-Diptera on C) are symbolized by arrows and superimposed on the individuals (nuclear receptors). The plots display, either the first and second principal axes (A), or the second and third principal axes (B). Eigenvalues bar charts show, in black, the two axes used to draw each biplot. Clustering dendrogram (D) based on the position of nuclear receptors on the factorial map 2-3 (B) was computed following Ward's method. Bold branches underline the clusters, which are found using four different hierarchical methods (see materials and methods). The supplementary points on A and B correspond to the projections of insect's phylogenomic trees obtained with concatenated alignments of large numbers of genes: ▲ 33,809 aa (Savard *et al.*, 2006b), ■ 336,069 aa and ● 705,502 aa (Zdobnov and Bork, 2007).

Figure 5. Principal component analysis of the evolution of transcription factors involved at the top of the ecdysone regulatory cascade. The same analysis as presented in figure 4 was performed with the eight nuclear receptors involved in the early ecdysone pathway plus E74, E93, Kr-h1 and BR (A). The PCA biplot (second and third principal axes) is built and reported with the same conventions as in figure 4. (B) A predefined phylogenetic tree is used to compute the branch lengths standing for variables in the PCA; the length of *Drosophila* branch (dotted line) was not retained for the PCA due to an atypical strong divergence of the *Drosophila* protein E93 (C) Clustering dendrogram based on the position of genes on the factorial map was computed following Ward's method. Bold branches underline the clusters, which are found using four different hierarchical methods (see materials and methods).

Figure 6. Summary of the ecdysone regulatory cascade, with the twelve transcription factors known to act as classic early regulators during the onset of *Drosophila* metamorphosis. After: Thummel (2001); King-Jones *et al.*, (2005); King-Jones and Thummel (2005). Nuclear receptors are boxed. The six proteins that overaccelerated in Mecopterida are in red. The known protein-protein interactions are indicated by large black bonds.

| NR nomenclature | Name[a] | Drosophila ortholog | Tribolium Accession | LG[b] | DBD/LBD %identity[c] |
|---|---|---|---|---|---|
| NR1D3 | E75 (REVERB) | Ecdysone-induced protein 75B CG8127 | TC_12440 | 9 | 99/58 |
| NR1E1 | E78 | Ecdysone-induced protein 78C CG18023 | TC_03935 | 3 | */60 |
| NR1F4 | HR3 (ROR) | Hormone receptor-like in 46 CG33183 | TC_08909 | 7 | 97/62 |
| NR1H1 | ECR (LXR/FXR) | Ecdysone receptor CG1765-PA CG1765-PB | TC_12112 Ecra :AM295015 TC_12113 Ecrb :AM295016 | 9 | 88/66 |
| NR1J1 | HR96 | Hormone receptor-like in 96 CG11783 | TC_10645 | ? | 78/53 |
| NR2A4 | HNF4 (HNF4) | Hepatocyte nuclear factor 4 CG9310 | TC_08726 | 7 | 94/78 |
| NR2B4 | USP (RXR) | Ultraspiracle CG4380 | TC_14027 TC_14028 AM295014 | 5 | 94/45 |
| NR2D1 | HR78 | Hormone-receptor-like in 78 CG7199 | TC_04598 | 1=X | 90/37 |
| NR2E2 | TLL (TLX) | Tailless CG1378 | TC_00441 AAF71999 | 2 | 81/38 |
| NR2E3 | HR51 (PNR) | Hr51 CG16801 | TC_09378 | 7 | 97/67 |
| NR2E4 | DSF | dissatisfaction CG9019 | TC_01069 TC_01070 | 2 | 95/68 |
| NR2E5 | HR83 | HR83 CG10296 | TC_10460 | ? | 76/26 |
| NR2E6[d] | *Nameless* | *No ortholog* | TC_13148 | 5 | * |
| NR2F3 | SVP (COUP-TF) | seven up CG11502 | TC_01722 | ? | 98/94 |
| NR3B4 | ERR (ERR) | estrogen-related receptor CG7404 | TC_09140 TC_09141 | 7 | */54 |
| NR4A4 | HR38 (NURR1) | Hormone receptor-like in 38 CG1864 | TC_13146 | 5 | 98/78 |
| NR5A3 | FTZ-F1 (SF1) | ftz transcription factor 1 CG4059 | TC_02550 | 3 | 98/74 |
| NR5B1 | HR39 | Hormone receptor-like in 39 CG8676 | TC_14986 | 6 | 90/77 |
| NR6A1 | HR4 (GCNF1) | Hr4 CG16902 | TC_00543 | 2 | 96/56 |
| NR0A1 | KNI | knirps CG4717 | *No ortholog* | * | * |
| NR0A2 | KNRL | knirps-like CG4761 | TC_03413 | 3 | 97/* |
| NR0A3 | EG | eagle CG7383 | TC_03409 CAF21851 | 3 | 93/* |

**Table 1.** Nuclear receptors of *Tribolium castaneum*.

[a]names used in this article ; the name of one clear vertebrate orthologue of the same group is given into brackets. [b]LG : Linkage Group. [c]DBD/LBD identity : % amino-acid identity between the homologous *Tribolium* and *Drosophila* proteins. [d]not an official nomenclature name ; proposed by Velarde *et al*. (2006). * : no data

| NR | | *Mecopterida* | | | Other insects | Outgroup | Aligned sites |
|---|---|---|---|---|---|---|---|
| | | Diptera | Lepidoptera | Bootstrap | | | |
| NR1D3 | E75 | 2 | 4 | 100/100 | 3 | E75 crustacea | 492 |
| NR1E1 | E78 | 3 | 0 | 92/57 | 2 | E75 arhtropoda[b] | 307 |
| NR1F4 | HR3 | 4 | 5 | 99/100 | 3 | ROR vertebrates | 406 |
| NR1H1 | ECR | 9 | 5 | 100/100 | 7 | ECR arthropoda[b] | 370 |
| NR1J1 | HR96 | 4 | 0 | 63/76 | 2 | NR1I vertebrates | 312 |
| NR2A4 | HNF4 | 4 | 2 | -/- | 2 | HNF4 vertebrates | 319 |
| NR2B4 | USP | 7 | 6 | 100/100 | 7 | USP arthropoda[b] | 298 |
| NR2D1 | HR78 | 4 | 1 | 98/82 | 3 | NR2C vertebrates | 275 |
| NR2E2 | TLL | 5 | 1 | -/- | 2 | TLL vertebrates | 335 |
| NR2E3 | HR51 | 4 | 0 | 96/83 | 2 | PNR vertebrates | 200 |
| NR2E4 | DSF | 4 | 0 | 75/- | 2 | TLL insects | 230 |
| NR2E5 | HR83 | 3 | 0 | 96/97 | 2 | NR2E3-6 insects | 201 |
| NR2F3 | SVP | 4 | 1 | 20/- | 2 | COUP-TF vertebrates | 300 |
| NR3B4 | ERR | 5 | 0 | 100/100 | 2 | ERR vertebrates | 265 |
| NR4A4 | HR38 | 4 | 1 | -/- | 2 | NR4A vertebrates | 341 |
| NR5A3 | FTZ-F1 | 4 | 2 | 80/- | 2 | FTZ-F1 crustacea | 427 |
| NR5B1 | HR39 | 3 | 1 | -/- | 2 | FTZ-F1 arthropoda | 342 |
| NR6A1 | HR4 | 4 | 3 | 95/98 | 3 | NR6A vertebrates | 359 |

**Table 2. Summary of insect's nuclear receptor phylogenies, with emphasis on the Mecopterida divergence.** The number of aligned proteins is indicated for each group of insect. The boostrap values (Neighbour Joining/Maximum Parsimony) associated to the Diptera or to the Mecopterida branch are given for each receptor. - : no Mecopterida branch.

| NR | Name | *Tribolium* | *Apis* | *Anopheles* | *Drosophila mel.* |
|---|---|---|---|---|---|
| 2D1 | HR78 | 1=X | 2.31 | 30E / 3R | 78d1 / 3L |
| 2E4 | DSF | 2 | 1.56 | 34C / 3R | 26a1 / 2L |
| 6A1 | HR4 | 2 | 3.31 | 20B / 2L | 2c1 / X |
| 2E2 | TLL | 2 | 4.15 | 5A / X | 100a3 / 3R |
| 5A3 | FTZ-F1 | 3 | 1.1 | 22A / 2L | 75d6-7 / 3L |
| 1E1 | E78 | 3 | 6.28 | 25D / 2L | 78c1-2 / 3L |
| 0A3 | EG | 3 | 10.6-10.8 | 38B / 3L | 78e5-6 / 3L |
| 0A2 | KNRL | 3 | 10.9 | 38B / 3L | 77c7-d1 / 3L |
| 0A1 | KNI | - | - | - | 77e1 / 3L |
| 2B4 | USP | 5 | 9.18 | 10D / 2R | 2c5 / X |
| 2E6 | NR2E6 | 5 | 12.30 | - | - |
| 4A4 | HR38 | 5 | ? | 30C / 3R | 38c9-10 / 2L |
| 5B1 | HR39 | 6 | 6.53 | 34A / 3R | 39a1-2 / 2L |
| 1F4 | HR3 | 7 | 1.55 | 32A / 3R | 46f5-6 / 2R |
| 2E3 | HR51 | 7 | 1.56 | 37D / 3R | 51f7 / 2R |
| 3B4 | ERR | 7 | 5.26 | 10D / 2R | 66b9 / 3L |
| 2A4 | HNF4 | 7 | 9.18 | 10D / 2R | 29e2 / 2L |
| 1H1 | ECR | 9 | 8.36 | 46D / 3L | 42a9-13 / 2R |
| 1D3 | E75 | 9 | 11.28 | 46D / 3L | 75a10-b6 / 3L |
| 2F3 | SVP | ? | 6.45 | 11B / 2R | 87b4 / 3R |
| 2E5 | HR83 | ? | 10.37 | 83e4 / 3R | 83e4 / 3R |
| 1J1 | HR96 | ? | 15.20 | 18C / 2R | 96b10-11 / 3R |

Supplementary Table. Chromosomal location of nuclear receptors in insects.

Figure 1

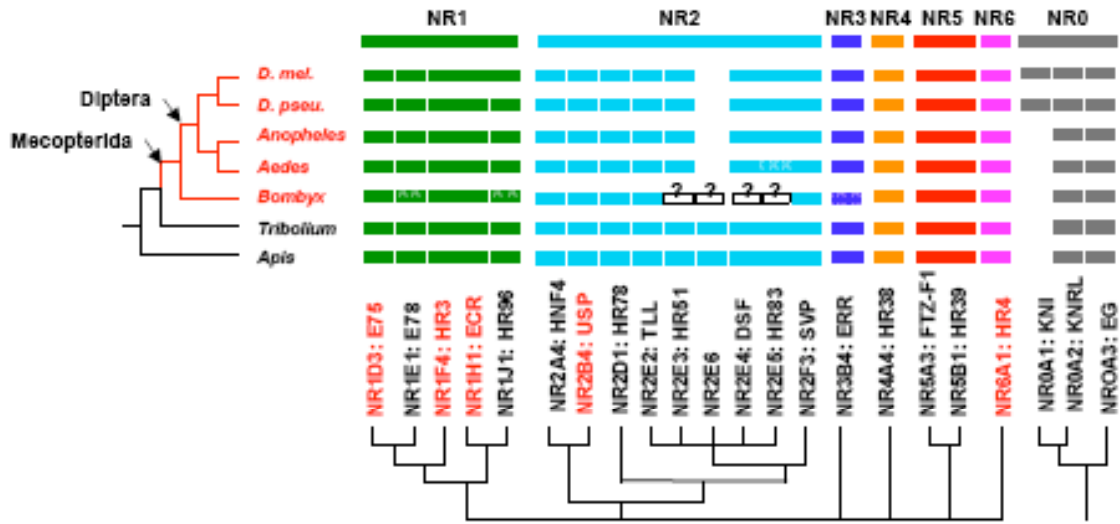

Figure 2A

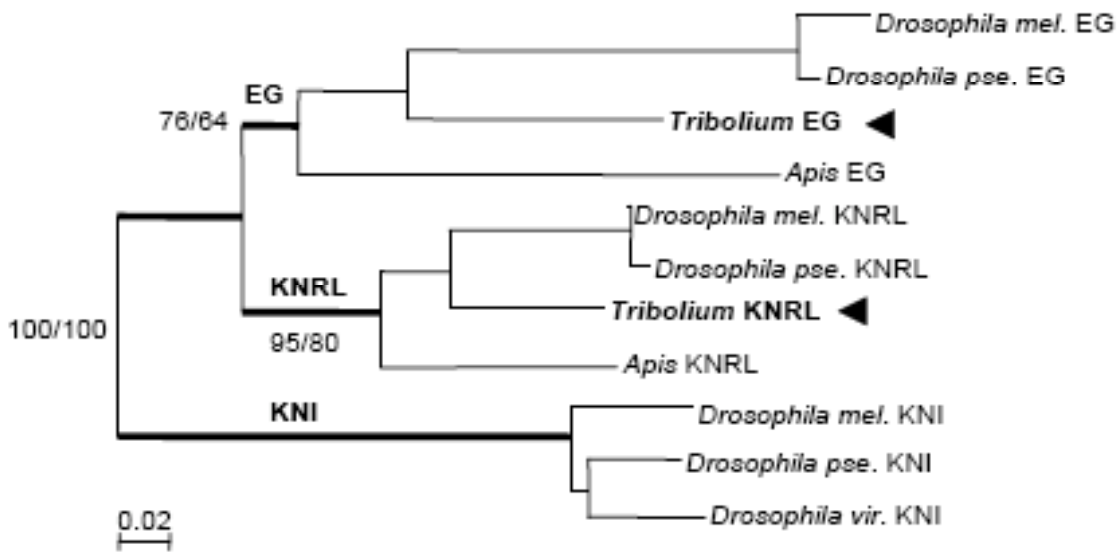

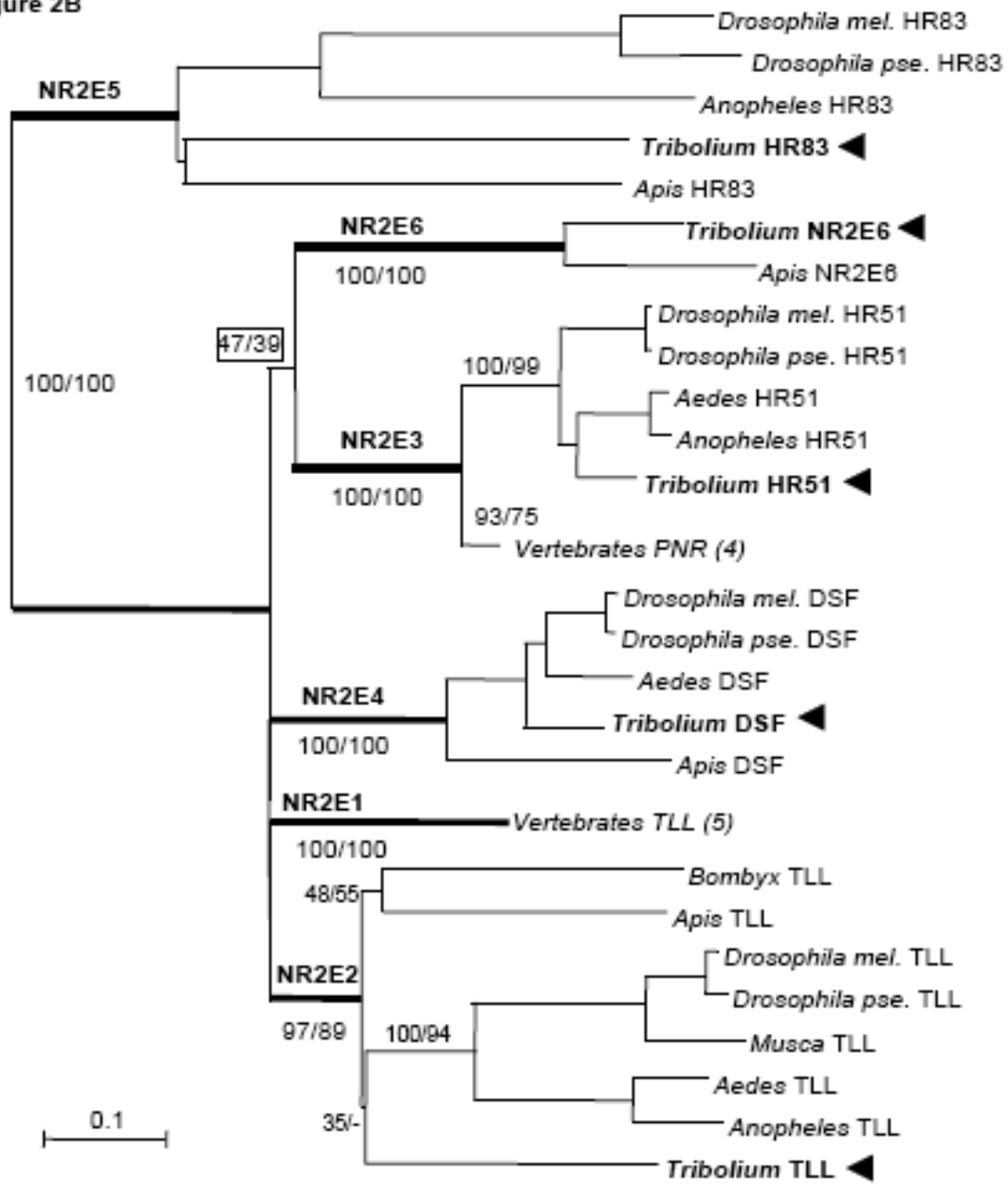

Figure 2B

**Figure 3**

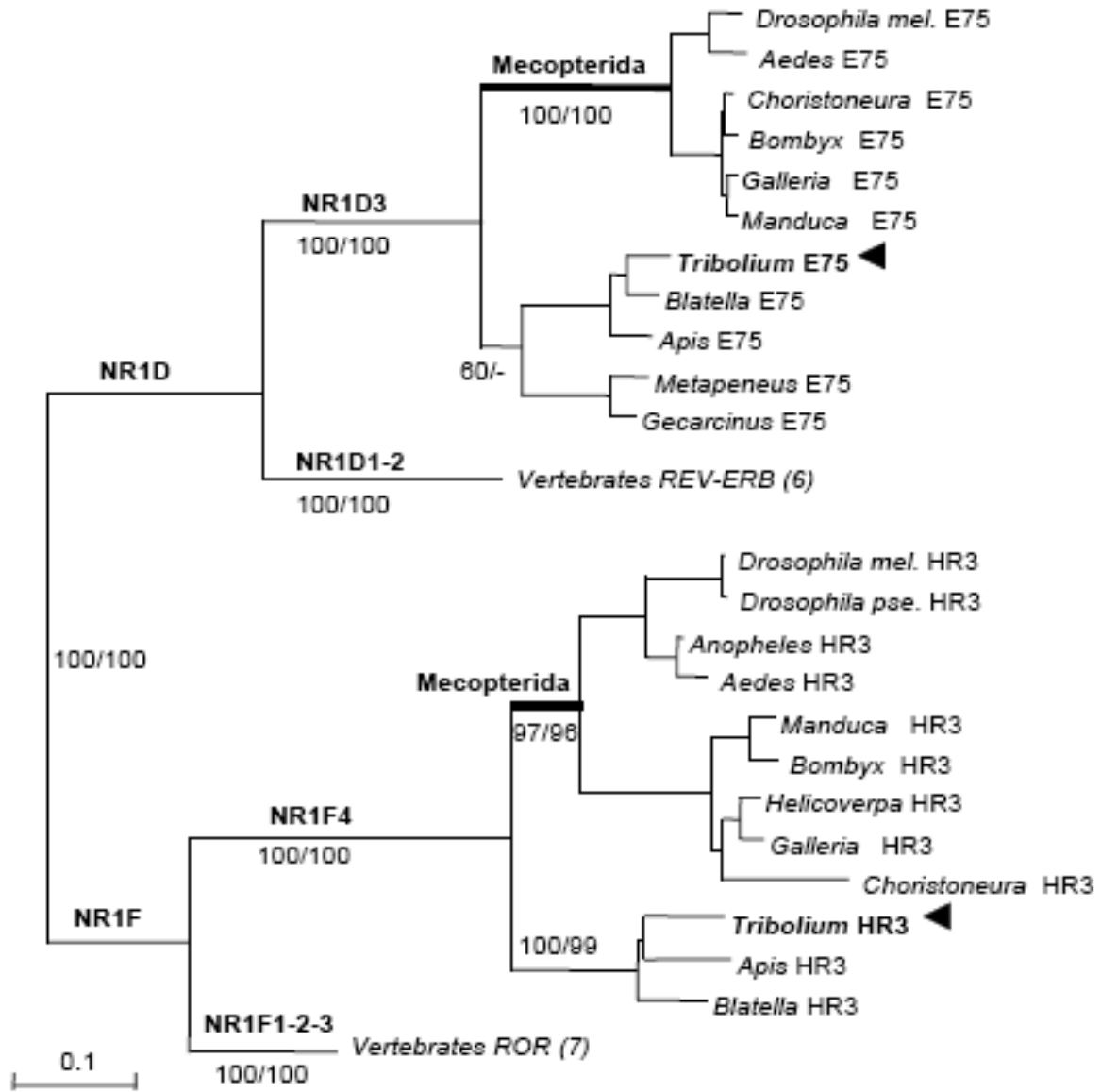

fig4

Figure 4

fig5

Figure 6

Late genes

Supplementary Figure 1

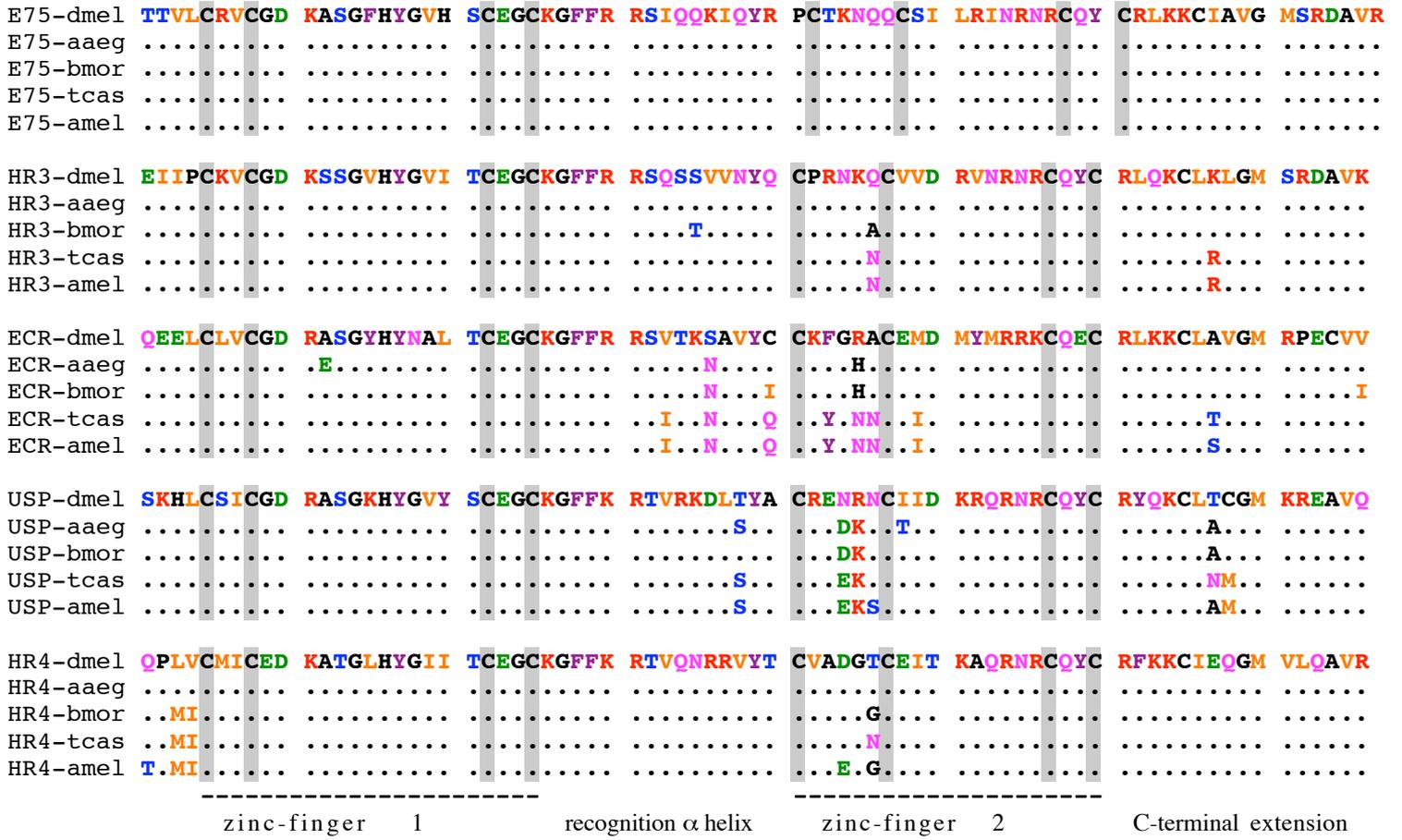

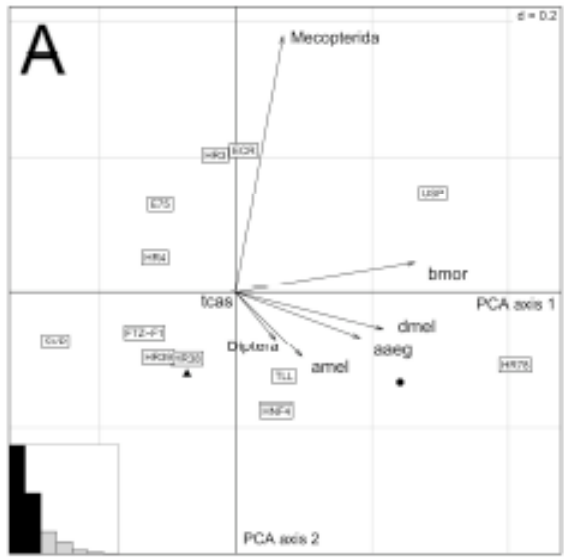
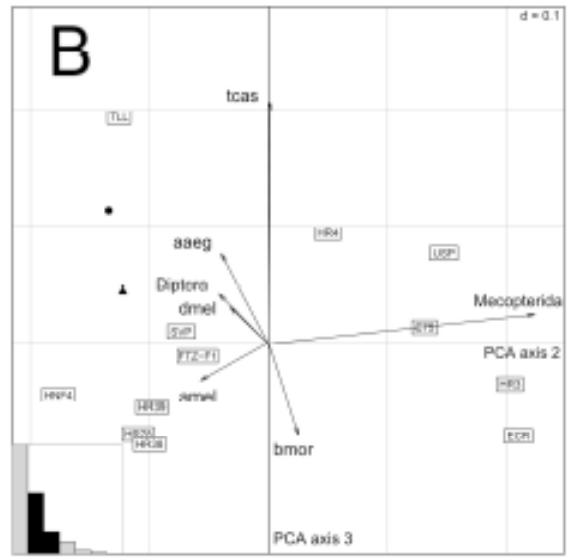
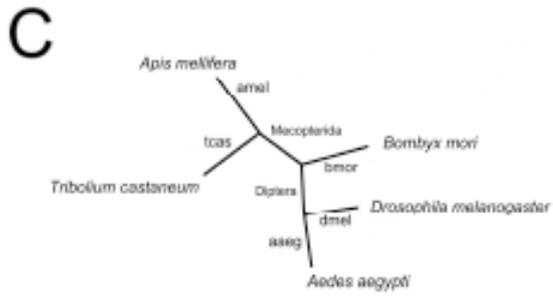
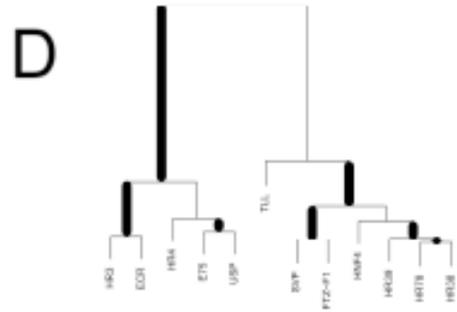